\documentclass[camera]{jpaper}

\usepackage[nocompress]{cite}
\usepackage{algorithmic}
\usepackage{array}

\makeatletter
\let\MYcaption\@makecaption
\makeatother

\usepackage[font=footnotesize]{subcaption}

\makeatletter
\let\@makecaption\MYcaption
\makeatother

\usepackage{fixltx2e}
\usepackage{dblfloatfix}
\usepackage[nolessnomore, italic]{mathastext}
\usepackage[T1]{fontenc}
\usepackage{textcomp}
\usepackage[usenames,dvipsnames,svgnames,table]{xcolor}
\usepackage[normalem]{ulem}
\usepackage{enumitem}
\usepackage{setspace}
\usepackage{indentfirst}
\usepackage{footmisc}
\usepackage{fancyhdr}
\usepackage{authblk}
\usepackage[us,12hr]{datetime}
\usepackage[keeplastbox]{flushend}
\usepackage[hidelinks]{hyperref}

\newboolean{publicversion}
\setboolean{publicversion}{true}

\definecolor{darkgreen}{RGB}{70,168,70}

\ifthenelse{\boolean{publicversion}}{
  \pagenumbering{arabic}
  \newcommand{\grumbler}[2]{}
  \newcommand{\assign}[1]{}
  \newcommand{\respond}[3]{}
  \newcommand{\changesI}[0]{}

}{
  \pagenumbering{arabic}
  \newcommand{\grumbler}[2]{\textcolor{blue}{\bf #1: #2}}
  \newcommand{\assign}[1]{\textcolor{purple}{\bf RESPONSIBLE: #1}}
  \newcommand{\respond}[3]{\textcolor{#1}{\bf #2-response: #3}}
  \newcommand{\changesI}[1]{\textcolor{BrickRed}{#1}}

}

\newif\ifcameraready
\camerareadytrue

\newcommand{\versionnum}[0]{4.2}

\ifcameraready
\else
\fi

\ifcameraready
  \newcommand{\todo}[1][]{}
\else
  \newcommand{\todo}[1][]{\textbf{\fcolorbox{black}{red}{\color{white}{TODO}}} \underline{$\overline{\hbox{\emph{#1}}}$}}
\fi

\begin{document}
%
\title{A Memory Controller with Row Buffer Locality Awareness\\for Hybrid Memory Systems}


\author{%
HanBin Yoon$^{1,2}$\qquad
Justin Meza$^{3,2}$\qquad
Rachata Ausavarungnirun$^2$%
\vspace{2pt}\\%
Rachael A. Harding$^{4,2}$\qquad
Onur Mutlu$^{5,2}$}
\affil{
{\em $^1$Google\qquad
$^2$Carnegie Mellon University\qquad
$^3$Facebook}%
\vspace{2pt}\\
{\em $^4$Massachusetts Institute of Technology  \qquad
$^5$ETH Z{\"u}rich}}


\maketitle

\begin{abstract}

This paper summarizes the idea and key contributions of the Dynamic Row Buffer Locality Aware Memory
Controller (RBLA), which was published in ICCD 2012~\cite{rbla}, and
examines the work's significance and future potential.
Non-volatile memory (NVM) is a class of promising \changesI{scalable} memory technologies that
can \changesI{potentially} offer higher capacity than DRAM \changesI{at the same cost point}.  Unfortunately, the access latency and
energy of NVM is often higher than \changesI{those of} DRAM, while the endurance of NVM is
lower.  Many DRAM--NVM \emph{hybrid memory systems}, also known as \emph{heterogeneous memory systems},
use DRAM as a cache to NVM, to
achieve the low access latency, low energy, and high endurance of DRAM, while
taking advantage of the large capacity of NVM.  A key question for a hybrid memory
system is what data to cache
in DRAM to best exploit the advantages of each technology while avoiding the \changesI{disadvantages of each technology}
as much as possible.


We propose a new memory controller \changesI{design} that improves hybrid memory
performance and energy efficiency. We observe that both DRAM
and NVM banks employ row buffers that act as a cache for the most
recently accessed memory row. Accesses that are row buffer hits incur
similar latencies (and energy consumption) in both DRAM and NVM, whereas
accesses that are row buffer misses incur longer latencies (and higher
energy consumption) in NVM than in DRAM.  To exploit this, we devise a policy that
caches \changesI{heavily-reused} data that frequently misses in the NVM row \changesI{buffers} into DRAM.
Our policy tracks the row buffer miss counts of recently-used
rows in NVM, and caches in DRAM the rows that are predicted to incur
frequent row buffer misses.  Our proposed policy also takes
into account the high write latencies of NVM, in addition to row
buffer locality \changesI{and more likely places the write-intensive pages
in DRAM instead of NVM}.



\changesI{We} evaluate our proposal
using a hybrid memory consisting of DRAM and phase-change memory (PCM), a
representative type of non-volatile memory. Compared to a
conventional DRAM--PCM hybrid memory system \changesI{that caches frequently-accessed
data in DRAM}, our row buffer locality-aware
hybrid memory system improves average system performance by 14\%, and average energy efficiency by
10\%, on data-intensive server and cloud workloads. Our proposed hybrid memory system
achieves a 31\% performance gain over an all-PCM memory system, and comes within
29\% of the performance of an all-DRAM memory system (not taking PCM's capacity
benefit into account) on our evaluated workloads.

\end{abstract}


\section{Introduction}
\label{sec:intro}

Multiprogrammed \changesI{and multithreaded} workloads on chip multiprocessors require large
amounts of main memory to support the working sets of many
concurrently-executing threads.  The demand for memory is increasing
rapidly, as the number of cores \changesI{or accelerators (collectively called \emph{agents})} on a chip continues to increase and
data-intensive applications become more widespread~\cite{bigdatabench,cloudbench,superfri,mutlu.imw13}.  \textit{Dynamic
  Random Access Memory} (DRAM) is used to compose main memory in
modern computers.  Though strides in DRAM manufacturing process technology have
enabled DRAM to scale to smaller feature sizes, and, thus, higher
densities (capacity per unit area), it is predicted that DRAM density
scaling will result in higher costs and lower reliability as the process technology feature size continues to
decrease~\cite{itrs10,ibm_2008,mandelman,chang-sigmetric16,kang14, mutlu.imw13,lee-hpca2013,superfri,kim.isca14,mutlu2017rowhammer}. Satisfying increasingly
higher memory demands with \changesI{exclusively DRAM will soon become too} expensive in
terms of both cost and energy.\footnote{We refer the reader to our prior works~\cite{atlas,tcm,kim.isca12,lee-hpca2013,lee.hpca15,hassan.hpca16,liu.isca12,chang.hpca14,chang.hpca16,seshadri.micro13,
chang.sigmetrics17,lee.sigmetrics17,seshadri.micro17,liu.isca13, chang-sigmetric16, hassan.hpca17,
kim.cal15, lee.pact15, lee.taco16, kim.isca14, patel.isca17, kim.hpca18} for a detailed background on DRAM.}

%

\subsection{Non-Volatile Memory}
\label{sec:pcm}

Emerging \emph{non-volatile memory} (NVM) technologies such as
\emph{phase-change memory} (PCM)~\cite{archpcm,pcmfuture,wong,BenLeeCACM2010,strictcache,hanbin2014taco,meza2013weed}, \emph{spin-transfer torque magnetic RAM} (STT-MRAM)~\cite{Architect_STTRAM, guo-isca2009, chang-hpca2013,naeimi.itj13},
\emph{resistive RAM} (ReRAM)~\cite{chua.tct71, strukov.nature08, ReRam}, and 3D XPoint~\cite{3DXpoint},
have shown promise for
future main memory system designs to meet the increasing memory capacity demands of
\emph{data-intensive} workloads. With projected scaling trends, NVM cells can be
manufactured more easily at smaller feature sizes than DRAM cells, achieving
high density and
capacity~\cite{archpcm,pcmfuture,Architect_STTRAM,ReRam,ibm_2008,JSSC2013,IEDM2010,IEDM2009,strictcache,ISCA2009_Pittsburgh,BenLeeCACM2010,wong,guo-isca2009, chang-hpca2013,naeimi.itj13,hanbin2014taco,meza2013weed}.
This is due to two reasons:
(1)~while a DRAM cell stores data in the form of charge, an NVM cell uses \changesI{resistive values
to represent the data, which is}
expected to scale to smaller feature sizes; and
(2)~unlike DRAM, several NVM devices use \emph{multi-level cell} technology, which stores
more than one bit of data per memory cell.

For example, PCM is a non-volatile memory technology that
stores data by varying the electrical resistance of a material known
as chalcogenide~\cite{wong,ibm_2008,archpcm}.  A PCM memory cell is programmed
by applying heat (via electrical current) to the chalcogenide and then
cooling it at different rates, depending on the data to be stored.
Rapid quenching places the chalcogenide into an amorphous state which
has high resistance, representing the bit value of `0' in single-level cell PCM, and slow cooling places
the chalcogenide into a crystalline state which has low resistance,
representing the bit value of `1' in single-level cell PCM.
Multi-level cell PCM can store multiple bits of data by
providing more than two distinguishable resistance levels for each cell, \changesI{
very similar to the MLC NAND flash technology that is prevalent in modern storage
systems~\cite{cai.procieee17, cai.procieee.arxiv17,
cai.date13, cai.date12, cai.hpca15, cai.dsn15, luo.jsac16, cai.sigmetrics14, cai.iccd12, cai.itj13,
cai.hpca17, cai.iccd12, cai.bookchapter.arxiv17,
hanbin2014taco,qureshi.isca2010}}.






However, NVM has a number of disadvantages. \changesI{Compared} to DRAM, NVM typically has a
longer access latency, higher write energy, and lower endurance~\cite{archpcm,strictcache}.  For example,
PCM's long cooling duration required to crystallize chalcogenide leads to
high PCM write latency, high read (sensing) latency, high read energy, and high
write energy compared to those of DRAM~\cite{meza.iccd12}.  
Furthermore, the repeated thermal expansions and contractions of
a PCM cell during programming lead to \emph{finite write endurance}, which is
estimated at $10^8$ writes, an issue not present in DRAM~\cite{archpcm}.

\subsection{Hybrid Memory Systems}
\label{sec:hybridmem}

Hybrid memory systems~\cite{strictcache, pdram, thynvm, pcmand3ddiestacking,
timber,yang-ubm, archdesignheteromemory, mlp_heterogenous_memory,gai2016smart,dulloor2016data,thermostat,pena2014toward,bock2016concurrent,critical_word} aim to combine the strengths of DRAM and emerging memory technologies
\changesI{(e.g., NVM, reduced-latency DRAM~\cite{micron-rldram3, sato-vlsic1998,lee-hpca2013}, reduced
reliability DRAM~\cite{strictcache,luo-dsn2014}). Many previous} DRAM-NVM hybrid memory system
designs employ DRAM as a small cache~\cite{strictcache} or write buffer~\cite{pdram, pcmand3ddiestacking} to NVM of large capacity.
In this work, we utilize PCM to provide increased overall memory capacity (which
leads to reduced page faults in the system), while the DRAM cache serves a large
portion of the memory requests at low latency and low energy with high
endurance. The combined effect increases overall system performance and energy
efficiency~\cite{strictcache}.
A key question in the design of a DRAM-PCM hybrid memory system is how to place
data between DRAM and PCM to best exploit the strengths of each
technology while avoiding their weaknesses as much as possible.

\subsection{Memory Device Architecture}
\label{sec:rowbuffers}

\begin{figure}[h!]
  \centering
    \includegraphics[height=1.8in]{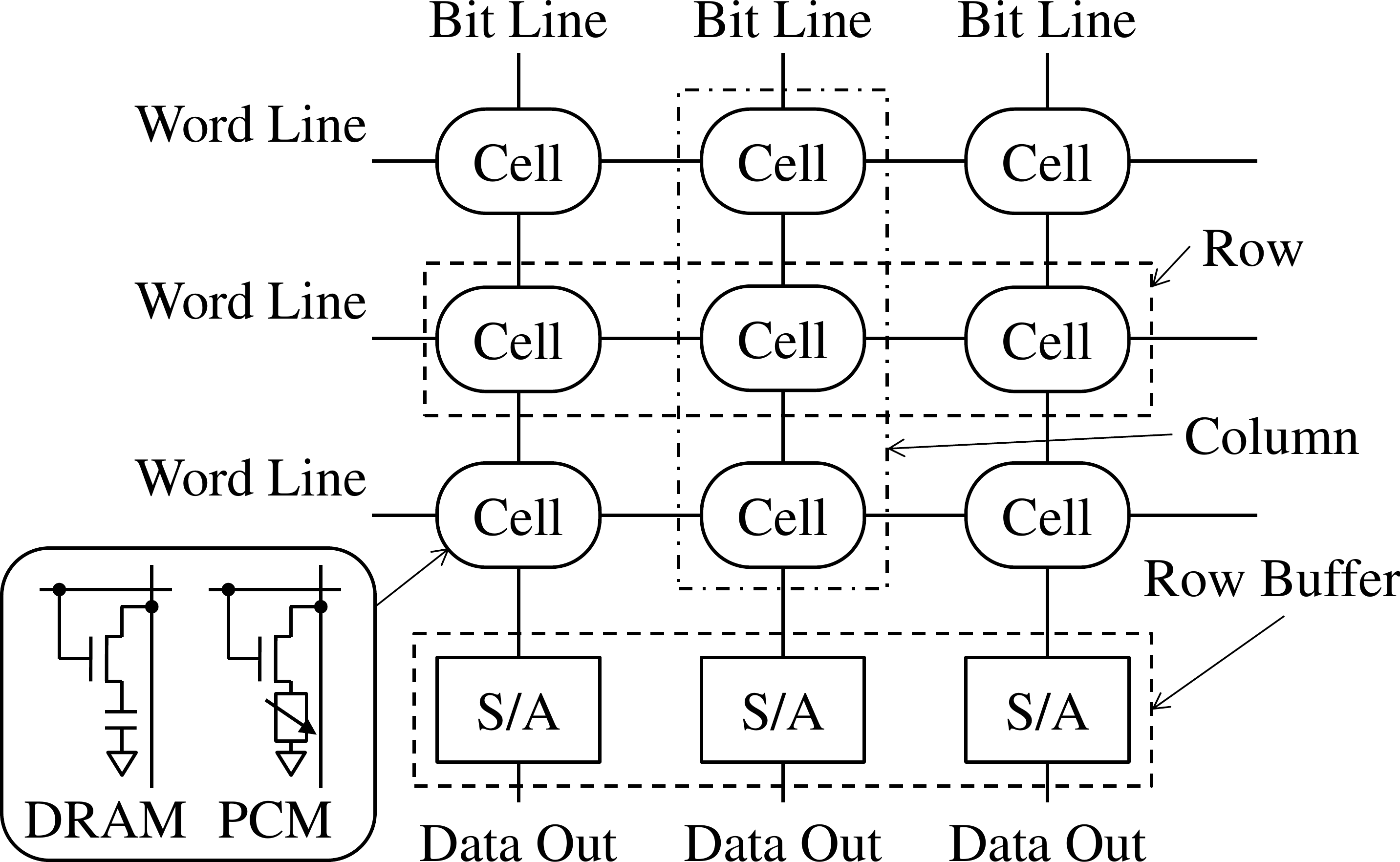}
    \caption{Memory cells organized in a 2D array of rows and columns. Reproduced from~\cite{rbla}.}
    \label{fig:memdevarch}
\end{figure}

In our ICCD 2012 paper~\cite{rbla},  we  develop  new  mechanisms  for  deciding
how data should be placed in a DRAM-PCM hybrid memory
system.  Our  main  observation  is  that  both  DRAM  and  PCM devices consist of
banks that employ  row  buffer  circuitry.  
The organization of a memory bank is illustrated in Figure~\ref{fig:memdevarch}.  
Cells (memory elements) are typically laid out
in arrays of rows (cells sharing a common \emph{wordline}) and columns
(cells sharing a common \emph{bitline}).  An access to the array occur at the
granularity of a row.  To read from the array, a wordline is first
asserted to select a row of cells.  Then, through the bitlines,
the contents of the selected cells are detected by sense amplifiers (labeled
S/A in the figure) and
latched by peripheral circuitry known as the \textit{row buffer}.

Once the contents of a row are latched in the row buffer, subsequent
memory requests to that row are served promptly from the row buffer,
without having to bear the delay of accessing the array.  Such memory
accesses are called \textit{row buffer hits}.  However, if a row
different from the one latched in the row buffer is requested, then
the newly requested row is read from the array into the row buffer
(replacing the row buffer's previous contents).  Such a memory access
incurs the high latency and energy of activating the array, and is
called a \textit{row buffer miss}.  \textit{Row buffer locality} (RBL)
refers to the repeated reference to a row while its contents are in
the row buffer.  Memory requests to data with high row buffer locality
are served efficiently (at low latency and energy) without having to
frequently \changesI{re-activate} the memory cell array.

\section{Row Buffer Locality-Aware Caching Policy}

\begin{figure}[b]
  \centering
    \includegraphics[height=1.8in]{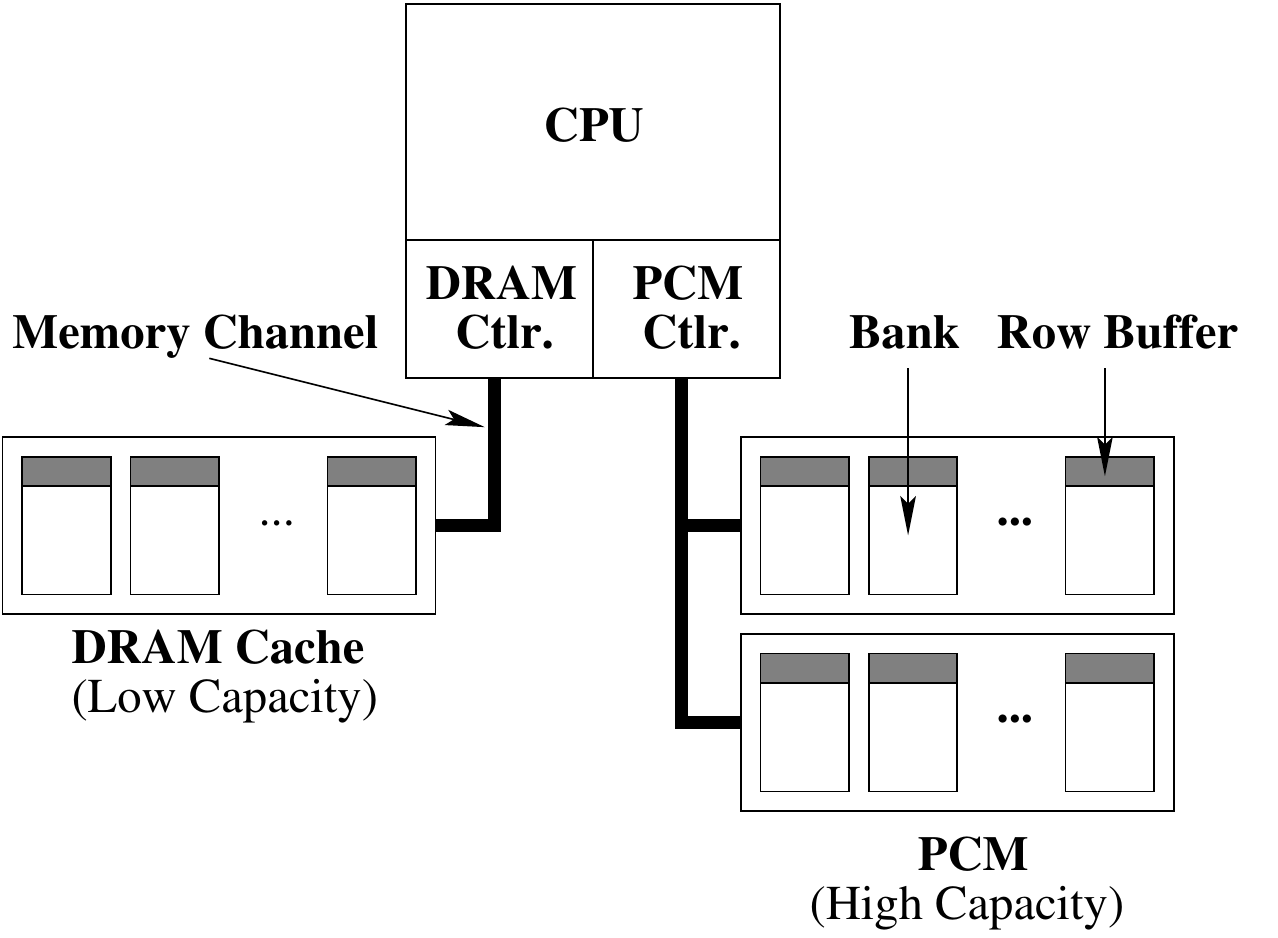}
    \caption{DRAM-PCM hybrid memory system organization. Reproduced from~\cite{rbla}.}
  \label{fig:org}
\end{figure}

\begin{figure*}[b]
  \centering
  \vspace{10pt}
  \includegraphics[width=\textwidth]{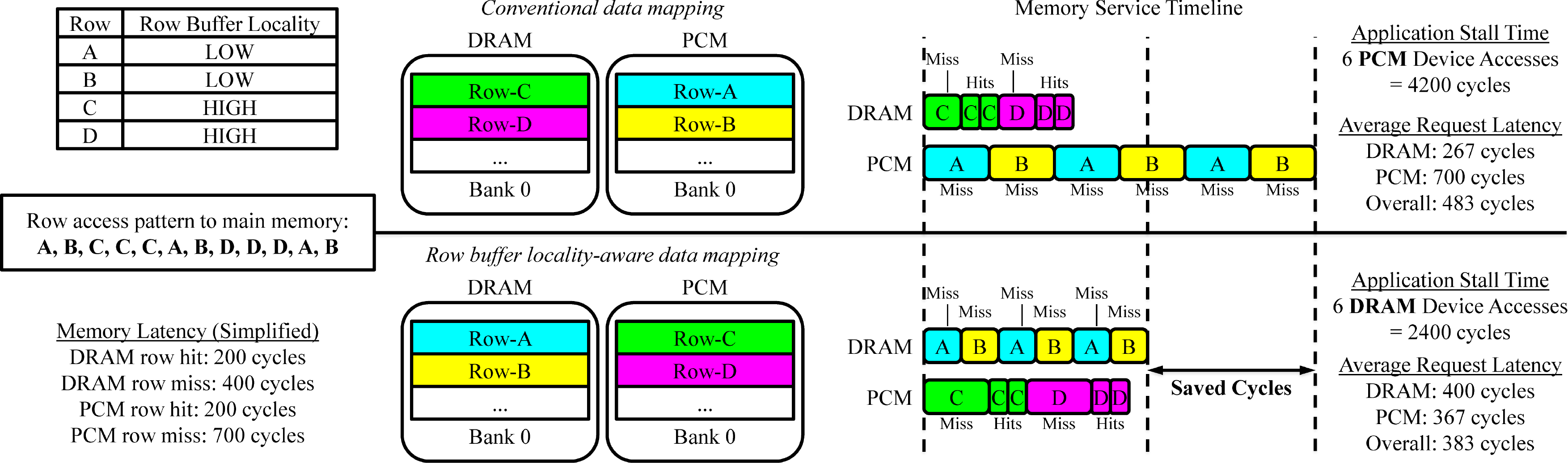}
  \caption{Conceptual example showing the importance of row buffer locality-awareness in hybrid memory data placement decisions. Reproduced from~\cite{rbla}.}
  \label{fig:access-stream}
\end{figure*}

Our ICCD 2012 paper~\cite{rbla} proposes \textit{Row Buffer Locality-Aware (RBLA)} caching policies, which a hybrid memory controller can use to guide data placement.
RBLA can be used in any hybrid memory system where each underlying memory
technology consists of banks with row buffers.
We study an example hybrid memory system that consists of a large amount of
PCM backed by a small DRAM cache~\cite{strictcache,missmap, timber,yang-ubm}, 
whose organization is shown in Figure~\ref{fig:org}.
Our main observation is that memory requests that \emph{hit} in the row buffer incur
\emph{similar} latencies and energy consumption in both DRAM and PCM~\cite{archpcm, pcmfuture}, whereas
requests that \emph{miss} in the row buffer incur \emph{higher} latency and energy in
PCM than in DRAM.
As a result, placing data that mostly leads to \emph{row buffer hits} (i.e., data that has high
row buffer locality) in DRAM provides \emph{little benefit} over placing the same
data in PCM. On the other hand, placing heavily reused data that leads to
frequent row buffer misses (i.e., data that has low row buffer locality) in DRAM avoids the
high latency and energy of PCM array accesses.

This observation is illustrated in the example shown in Figure~\ref{fig:access-stream}.
In the example, the service timelines for memory requests to rows A--D are shown.
Prior hybrid memory and cache management proposals seek to improve the
reuse of data placed in the cache and reduce the access bandwidth of the next
level of memory (e.g., \cite{pageplacement, chop}).  We call this approach to
cache management \emph{conventional mapping}.
Conventional mapping (top half of Figure~\ref{fig:access-stream}) can place rows A 
and B \changesI{(which have low row buffer locality) \emph{both} in PCM},
causing the high PCM array latency to become a bottleneck. In contrast, \textit{row
buffer locality-aware} mapping (bottom half of Figure~\ref{fig:access-stream}) places 
rows A and B in DRAM \changesI{such that they can} benefit
from DRAM's lower array latency, leading to faster overall memory service.\footnote{Even
though the figure shows some requests being served in parallel, if
the individual requests arrived in the same order at different times, the average
request latency would still be improved significantly.}
Placing rows C and D (high row locality) in DRAM provides little benefit over
placing them in PCM.

Based on this observation, we devise a 
hybrid memory caching policy that caches \changesI{in} DRAM the rows that mostly miss in
the row buffer and are frequently reused.
To implement this policy, the memory controller maintains a count of the
row buffer misses for \changesI{recently-used} rows in PCM, and places in DRAM
the data of rows whose row buffer miss counts exceed a certain
threshold (dynamically adjusted at runtime in the RBLA-Dyn mechanism, which we 
describe in Section~\ref{sec:dinner-blah}).


\subsection{Measuring Row Buffer Locality}

The RBLA mechanism tracks the row buffer locality statistics for a small
number of recently-accessed rows, in a hardware structure called the
\textit{stats store}. The stats store resides in the memory controller,
and is organized similarly to a cache, however its data payload per entry
is a single row buffer miss counter.

On each PCM access, the memory controller looks for an entry in the
stats store using the address of the accessed row.  If there is no
corresponding entry, a new entry is allocated for the accessed row,
possibly evicting an older entry.  If the access results in a row
buffer miss, the row's row buffer miss counter is
incremented. 
If the access results in a
row buffer hit, no additional action is taken.

\subsection{Triggering Row Caching}

Rows that exhibit low row buffer locality and high reuse will have high
row buffer miss counter values.  The RBLA mechanism selectively caches
these rows by \changesI{triggering the caching of a row in} DRAM when 
\changesI{the row's} row buffer miss counter
exceeds a threshold value, \texttt{MissThresh}.  Setting this
\texttt{MissThresh} to \changesI{a low value} causes more rows with a higher row buffer
locality to be cached.

Caching rows based on their row buffer locality attempts to migrate
data between PCM and DRAM only when such data movement is beneficial. This
affects system performance in three ways.  First, placing in DRAM
rows that have low row buffer locality improves average memory access
latency, due to the lower row buffer miss latency of
DRAM compared to PCM.  Second, by selectively caching data that
benefits from being migrated to DRAM, RBLA reduces unnecessary data
movement between DRAM and PCM (i.e., data that frequently hits in the row buffer
incurs the same access latency in PCM as in DRAM, and is thus left in PCM).
This reduces memory bandwidth consumption, allowing more bandwidth to be used to serve
demand requests, and enables better utilization of the DRAM cache space.
Third, allowing data that frequently hits in the row buffer to remain
in PCM contributes to balancing the memory request load between DRAM and PCM.

To prevent rows with low reuse from gradually building up large enough
row buffer miss counts over an extended period of time to exceed
\texttt{MissThresh} and trigger row caching, we apply a periodic reset
to all of the row buffer miss count values.  We set this reset interval
to 10 million cycles empirically.

\subsection{Dynamic Threshold Adaptation: RBLA-Dyn}
\label{sec:dinner-blah}

We improve the adaptivity of RBLA to workload and system variations
by \emph{dynamically} determining the value of \texttt{MissThresh}.
The key idea behind this scheme, which we call RBLA-Dyn, is that \textit{the number of
  cycles saved by caching rows in DRAM should outweigh the
  cost of migrating that data to DRAM}.  RBLA-Dyn estimates, on an
interval basis, the first order cost and benefit of employing
a certain \texttt{MissThresh} value, and increases or decreases the
\texttt{MissThresh} value to maximize the \emph{net benefit} (i.e., benefit minus cost).

Since data migration operations can delay demand requests, we approximate cost
as the number of cycles spent migrating each row across the memory
channels ($t_{migration}$) times the number of rows migrated ($NumMigrations$):
\begin{equation}
\mathit{Cost}    = \mathit{NumMigrations} \times t_{\mathit{migration}} \label{eqn:cost}
\end{equation}
If these data migrations are
eventually beneficial, the access latency to main memory will decrease.
Hence, we can compute the benefit of migration as the number of cycles saved by
accessing the data from the DRAM cache as opposed to PCM:
\begin{align}
\mathit{Benefit} = & \mathit{NumReads}_{\mathit{dram}} \times (t_{\mathit{read},\mathit{pcm}} - t_{\mathit{read},\mathit{dram}}) + \label{eqn:benefit} \\
                   & \mathit{NumWrites}_{\mathit{dram}} \times (t_{\mathit{write},\mathit{pcm}} - t_{\mathit{write},\mathit{dram}}) \nonumber
\end{align}
In this equation, $\mathit{NumReads}_{\mathit{dram}}$ and $\mathit{NumWrites}_{\mathit{dram}}$
are the number of reads and writes performed in DRAM after migration, 
$t_{\mathit{read},\mathit{dram}}$ and $t_{\mathit{write},\mathit{dram}}$ are the
read and write latency of a DRAM row buffer miss, and 
$t_{\mathit{read},\mathit{pcm}}$ and $t_{\mathit{write},\mathit{pcm}}$ are the
read and write latency of a PCM row buffer miss.  RBLA-Dyn accounts for reads and writes
separately, as they incur different latencies in many NVM technologies, such as PCM.

RBLA-Dyn uses a simple hill-climbing algorithm (see Algorithm 1 in our ICCD 2012 paper~\cite{rbla})
to find the value of \texttt{MissThresh} that maximizes the net benefit.
The algorithm is executed at the end of each interval (10 million cycles in our setup).
We refer the reader to Section~IV-C of our ICCD 2012 paper~\cite{rbla} for more 
details on the RBLA-Dyn mechanism.


\subsection{Implementation and Hardware Cost}

The primary hardware cost incurred in implementing a row buffer
locality-aware caching mechanism on top of an existing hybrid memory
system is the stats store.  We model a 16-way, 128-set, LRU-replacement
stats store using 5-bit row buffer miss counters, which occupies
a total of 9.25~KB. 
This stats store
achieves within 0.3\% of the system performance (and within 2.5\% of the memory lifetime) 
of an unlimited-sized stats store for RBLA-Dyn.

\section{Evaluation Methodology}


%

We use a cycle-level in-house x86 multi-core simulator, whose
front-end is based on Pin. \changesI{The simulator is an early predecessor
of Ramulator~\cite{kim.cal15,ramulator} and the ThyNVM simulator~\cite{thynvm}.}
We collect results using multiprogrammed workloads consisting of server- and cloud-type 
applications (including TPC-C/H~\cite{tpc}, Apache
Web Server, and video processing benchmarks) for a 16-core system.
We compare our row buffer locality-aware caching policy (RBLA) against
a policy that caches data that is frequently accessed (FREQ, similar in approach to \cite{chop}).
We use this competitive baseline because we find that conventional LRU caching performs worse due
to its high memory bandwidth consumption.
FREQ caches a row when the number of accesses to the row exceeds a
threshold value.
FREQ-Dyn adopts the same dynamic threshold adjustment algorithm as RBLA-Dyn \changesI{(Section~\ref{sec:dinner-blah})}.
Our methodology and workloads are described in detail in Section~VI of our ICCD 2012 paper~\cite{rbla}.

\section{Evaluation}

\begin{figure}[h!]
  \centering
    \includegraphics[height=1.5in]{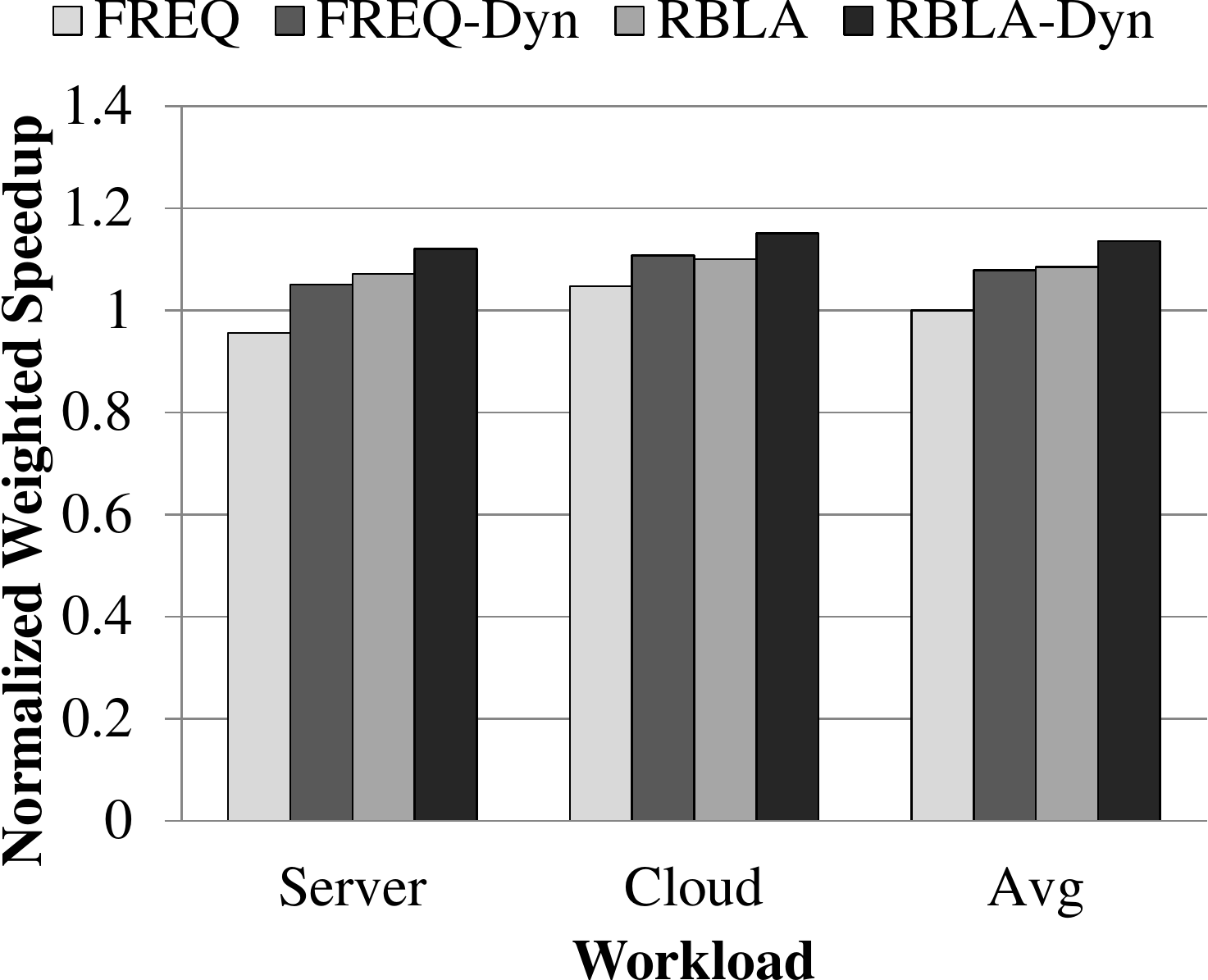}
  \caption{Weighted speedup of the four caching techniques: FREQ, FREQ-Dyn, RBLA, and RBLA-Dyn. Reproduced from~\cite{rbla}.}
    \label{fig:ws}
\end{figure}

\textbf{Performance.}
Figure~\ref{fig:ws} shows the weighted speedup of the four caching techniques
that we evaluate.
As we observe from the figure, RBLA-Dyn provides
the highest performance (14\% improvement in weighted speedup over FREQ) among 
the four techniques.
RBLA and RBLA-Dyn outperform FREQ and FREQ-Dyn, respectively, because
the RBLA techniques place data with low row buffer locality in DRAM where it can be
accessed at the lower DRAM array access latency, while keeping data
with high row buffer locality in PCM where it can be accessed at \changesI{the already-low} 
row buffer hit latency.

\begin{figure}[h!]
  \centering
    \includegraphics[height=1.5in]{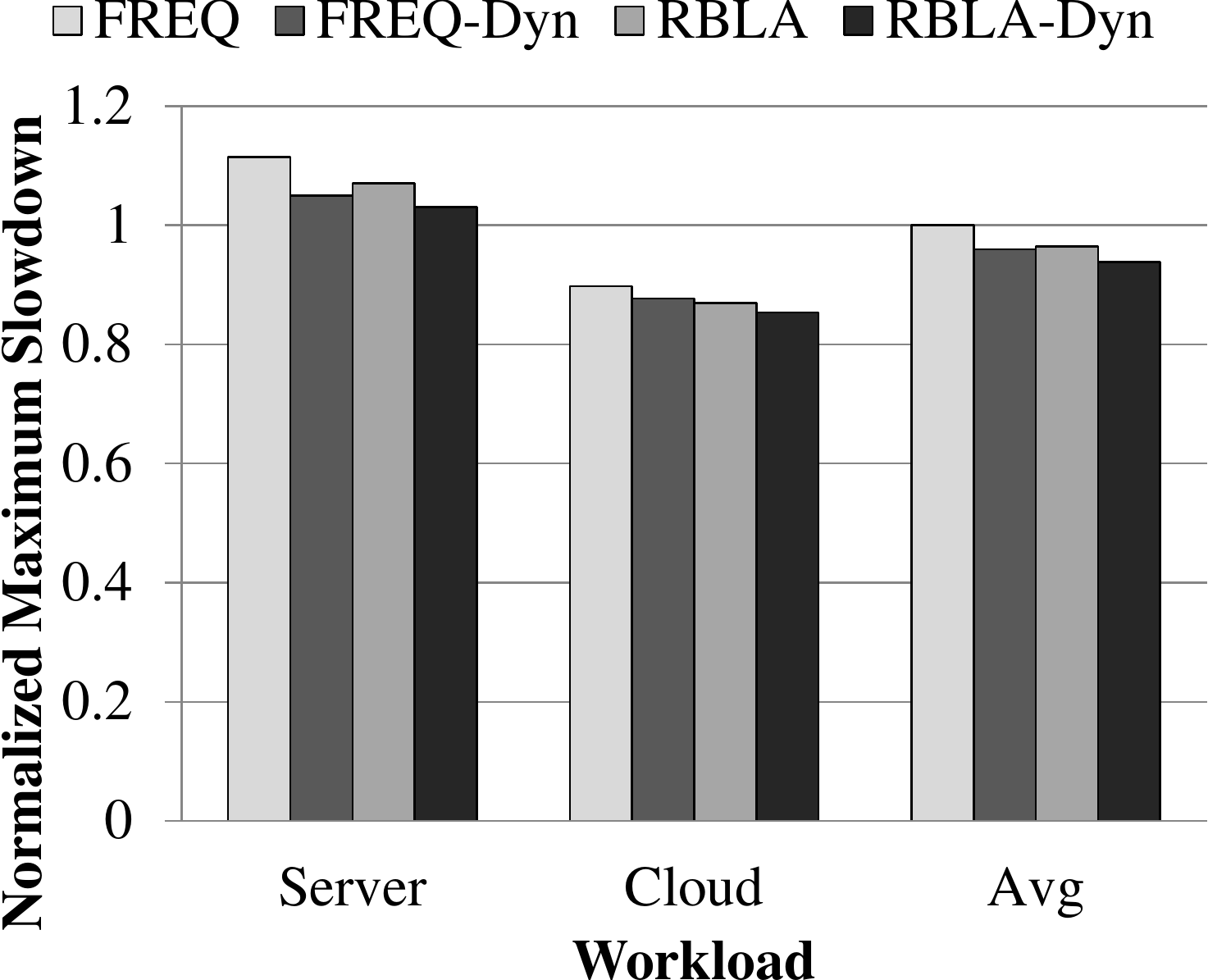}
  \caption{Fairness of the four caching techniques: FREQ, FREQ-Dyn, RBLA, and RBLA-Dyn (lower is better). Reproduced from~\cite{rbla}.}
    \label{fig:ms}
\end{figure}

\textbf{Thread Fairness.} 
Figure~\ref{fig:ms} shows the fairness of each caching technique.
We measure fairness using \emph{maximum slowdown}~\cite{Reetu-MICRO2009,atlas,tcm,sms,a2c,mcp,mise,usui-dash,vandierendonck,bliss,bliss-tpds}, which is
the highest slowdown (reciprocal of
speedup) experienced by any benchmark within the multiprogrammed
workload.  A \emph{lower} maximum slowdown indicates \changesI{\emph{higher} fairness}.
We observe from the figure that RBLA-Dyn provides the highest thread fairness (6\%
improvement in maximum slowdown over FREQ) out of all evaluated policies.  
RBLA-Dyn throttles back on
non-beneficial data migrations, reducing the amount of memory bandwidth and DRAM
space consumed due to such migrations.  Combined with the reduced average
memory access latency, RBLA-Dyn reduces contention for memory bandwidth among
co-running applications, providing \changesI{higher} fairness.

\begin{figure}[h!]
  \centering
    \includegraphics[height=1.5in]{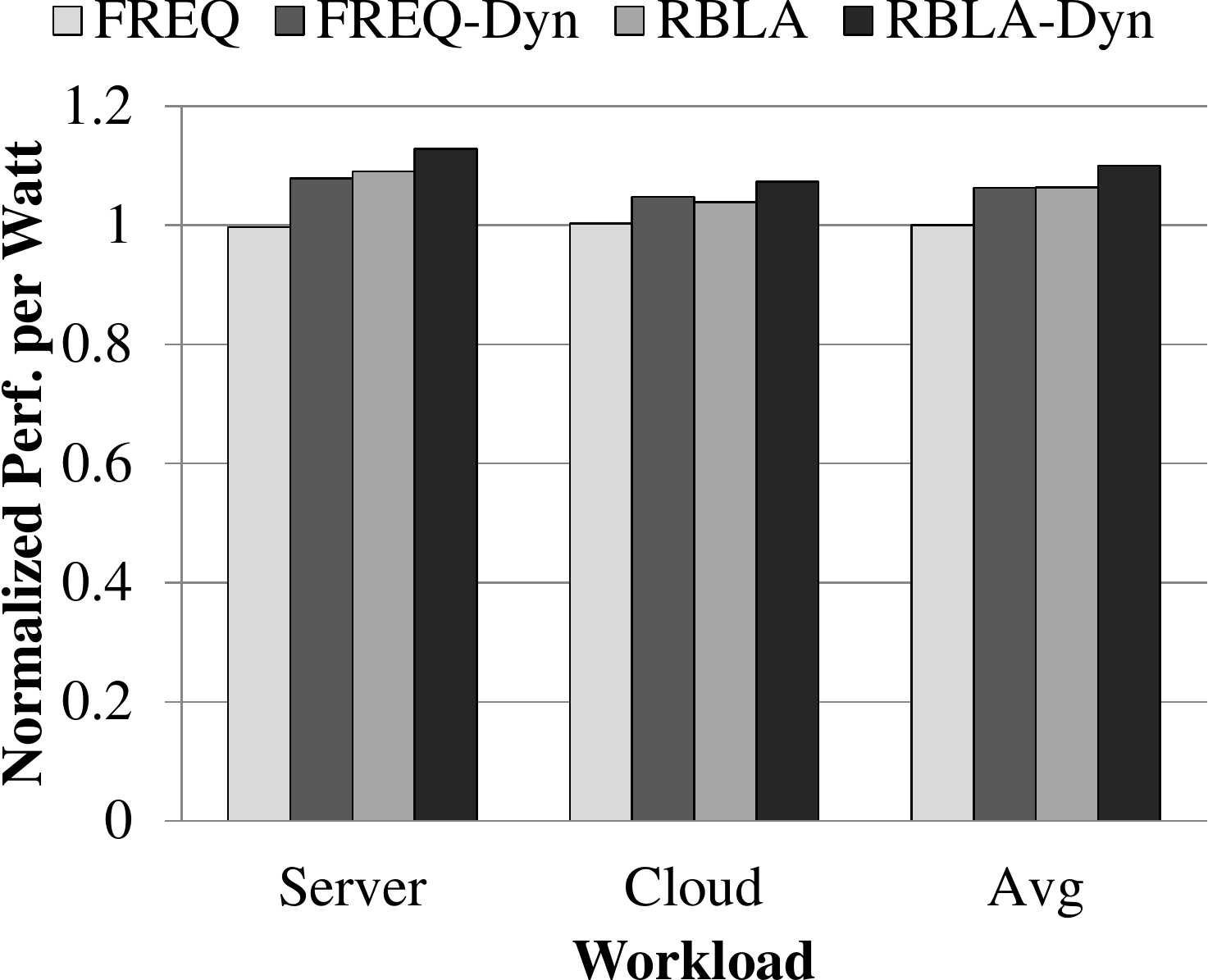}
  \caption{Energy efficiency of the four caching techniques: FREQ, FREQ-Dyn, RBLA, and RBLA-Dyn. Reproduced from~\cite{rbla}.}
    \label{fig:ee}
\end{figure}

\textbf{Memory Energy Efficiency.}
Figure~\ref{fig:ee} shows that RBLA-Dyn achieves the highest memory energy
efficiency (10\% improvement over FREQ) compared to other policies, in terms of
performance per Watt.  This is because RBLA-Dyn places data with low row buffer
locality in DRAM, making the energy cost of row buffer miss accesses lower than
it would be if such data were placed in PCM.  RBLA-Dyn also reduces energy
consumption by reducing the amount of non-beneficial \changesI{or useless} data migrations.

We provide the following other evaluation results in Section~VII of our ICCD 2012 paper~\cite{rbla}:
\begin{itemize}
\item Impact of RBLA-Dyn on average memory latency (Section~VII-A of \cite{rbla}).
\item Impact of RBLA-Dyn on DRAM and PCM channel utilization (Section~VII-A of \cite{rbla}).
\item Memory access breakdown \changesI{of each workload} to DRAM and PCM (Section~VII-A of \cite{rbla}).
\item Impact of RBLA-Dyn on PCM lifetime (Section~VII-D of \cite{rbla}).
\item Comparison with all-PCM and all-DRAM systems (Section~VII-E of \cite{rbla}).
\end{itemize}
As we discuss in detail in our ICCD 2012 paper~\cite{rbla},
RBLA-Dyn bridges the gap in performance between homogeneous all-DRAM and
all-PCM memory systems of equal addressable capacity (achieving within 29\%
of the performance of an all-DRAM system, and improving performance by 31\%
over an all-PCM system), while providing close to seven years of memory 
lifetime.\footnote{\changesI{Note that lifetime can be further improved by enabling more
aggressive write optimization~\cite{dbi}, and by taking advantage of 
application-level error tolerance~\cite{luo-dsn2014}.}}

We conclude that taking row buffer locality into account \changesI{enables new} hybrid
memory caching policies that achieve high performance and energy efficiency.

\section{Related Work}
\label{sec:related}

To our knowledge, our ICCD 2012 paper~\cite{rbla} is the first work to observe that row buffer
hit latencies are similar in different memory technologies, and uses
this observation to devise a caching policy that improves the
performance and energy efficiency of a hybrid memory system.  No
previous work, as far as we know, considered row buffer locality as a
key metric for deciding what data to cache and what not to cache.
We discuss related work on caching policies and hybrid memory systems.

\textbf{Caching Based on Data Access Frequency.}
Jiang et al.~\cite{chop} propose caching only the data that
experiences a high number of accesses in an on-chip DRAM cache (in
4--8~KB block sizes), to reduce off-chip memory bandwidth consumption.
Johnson and Hwu~\cite{johnson} use a counter-based mechanism to track
data reuse at a granularity larger than a cache block.  Cache blocks
in a region with less reuse bypass a direct-mapped cache if that
region conflicts with another that has more reuse.  We propose to take
advantage of row buffer locality in memory banks when employing
off-chip DRAM and PCM.  We exploit the fact that accesses to DRAM and
PCM have similar average latencies for rows that have high row buffer
locality.

Ramos et al.~\cite{pageplacement} adapt a buffer cache replacement
algorithm to rank pages based on their frequency and recency of
accesses, and place the highly-ranking pages in DRAM, in a DRAM-PCM
hybrid memory system.  Our work is orthogonal, because the page-ranking
algorithm can be adapted to rank pages based on their frequency and
recency of row buffer misses (not counting accesses that are row
buffer hits), for which we expect improved performance.








\textbf{Caching Based on Locality of Data Access.}
Gonzalez et al.~\cite{gonzalez} propose placing data in one of two
last-level caches depending on whether it exhibits spatial or temporal
locality.  They also propose bypassing the cache when accessing large
data structures with large strides (e.g., big matrices) to prevent
cache thrashing.
Rivers and Davidson~\cite{rivers} propose separating out data without
temporal locality from data with, and placing it in a special buffer
to prevent the pollution of the L1 cache.  These works are primarily
concerned with on-chip L1/L2 caches that have access latencies on the
order of a few to tens of processor clock cycles, where off-chip
memory bank row buffer locality is less applicable.

There have been many works in on-chip caching to improve cache
utilization (e.g., a recent one uses an evicted address filter to
predict cache block reuse~\cite{eaf-pact12}), but none of these
consider the row buffer locality of cache misses.

\textbf{Caching Based on Other Criteria.}
Chatterjee et al.~\cite{critical_word} observe that the first word of cache blocks
is critical to performance, and propose to store only the first word of each block in fast DRAM. 
Phadke and Narayanasamy~\cite{mlp_heterogenous_memory}
propose to classify applications into three categories  based on memory-level parallelism (MLP): 
latency-sensitive, bandwidth-sensitive,
and insensitive-to-both.
To estimate MLP, they profile the misses per kilo-instruction (MPKI) and stall time of each 
application offline during the compilation stage.
Applications with high MPKI but low stall time are considered to have good MLP.



\textbf{Hybrid Memory Systems.}
Qureshi et al.~\cite{strictcache} propose increasing the size of main
memory by adopting PCM, and using DRAM as a conventional cache to PCM.
The reduction in page faults due to the increase in main memory size
brings performance and energy improvements to the system. 
Our ICCD 2012 paper~\cite{rbla} proposes
a new, effective DRAM caching policy to PCM, and studies performance
effects without page faults present.

Li et al.~\cite{yang-ubm} propose UHM, a utility-based hybrid memory
management mechanism that expands upon our RBLA policy.
UHM estimates the \emph{utility} of each page, which is the
benefit to system performance of placing each page in different
types of memory (e.g., DRAM and NVM).
UHM migrates to the fast memory of a hybrid memory system only those pages 
whose utility would improve the most after migration.

\changesI{Ren et al.~\cite{thynvm} propose ThyNVM, which manages the DRAM and
PCM spaces carefully and adapts the granularity of management to the access
patterns in a manner that provides crash consistency in a 
persistent memory system.}

Dhiman et al.~\cite{pdram} propose a hybrid main memory system that
exposes DRAM and PCM addressability to the software (OS).  If the
number of writes to a particular PCM page exceeds a certain threshold,
the contents of the page are copied to another page (either in DRAM or
PCM), thus facilitating PCM wear-leveling.  Mogul et
al.~\cite{ossupportfornvmdram} suggest that the OS exploit metadata
information available to it to make data placement decisions between
DRAM and non-volatile memory.  Similar to \cite{pdram}, their data
placement criteria are centered around the write frequency to
data. Our proposal is complementary to this work, and row buffer locality information, if
exposed, can be used by the OS to place pages in DRAM or PCM.

Bivens et al.~\cite{archdesignheteromemory} examine the various
design concerns of a heterogeneous memory system such as memory
latency, bandwidth, and endurance requirements of employing storage
class memory (e.g., PCM, STT-MRAM, NAND flash memory).  Their hybrid memory
organization is similar to ours and that in \cite{strictcache}, in
that DRAM is used as a cache to a slower memory medium, transparently
to software.  Phadke et al.~\cite{mlp_heterogenous_memory} propose to profile the
memory access patterns of individual applications in a multi-core
system, and place their working sets in the particular type of DRAM
that best suits the application's memory demands.  In contrast, 
RBLA dynamically makes fine-grained data placement decisions at a
row granularity, depending on the row buffer locality characteristics \changesI{of each page}.

Agarwal et al.~\cite{thermostat} propose a software-based approach to manage huge pages (e.g., 2MB pages) 
in hybrid memory systems. The mechanism profiles the memory access patterns of huge pages,
and uses the profiling information to guide page migration between DRAM and NVM. 
Pe{\~n}a and Balaji~\cite{pena2014toward} propose a profiling tool to assess the impact of distributing 
memory objects across memory devices in hybrid memory systems. 
Bock et al.~\cite{bock2016concurrent} propose a scheme that allows concurrent migration of multiple pages between 
different types of memory devices without significantly affecting the memory bandwidth. 
Gai et al.~\cite{gai2016smart} propose a data placement scheme that aims to minimize the 
energy consumption of hybrid memory systems. Liu et al.~\cite{liu2016memos} propose 
a scheme that manages the entire memory hierarchy, which includes caches, memory channels, and DRAM/NVM banks. 
Dulloor et al.~\cite{dulloor2016data} propose a programmer-guided data
placement tool, which requires programmers to modify the source code, 
\changesI{and needs data from a 
representative profiling run of the application, prior to making}
placement decisions. 
Ideas from all of these works can be combined with RBLA for
better performance and efficiency.

\textbf{Exploiting Row Buffer Locality.}
Many previous works exploit row buffer locality to improve memory
system performance, but none (to our knowledge) develop a cache data
placement policy that considers the row buffer locality of the block
to be cached.
Lee et al.~\cite{archpcm, BenLeeCACM2010, pcmfuture} propose to use multiple short row buffers in
PCM devices, much like an internal device cache, to increase the row
buffer hit rate. \changesI{Meza et al.~\cite{meza.iccd12} examine the case for small row 
buffers for NVM devices}.
Sudan et al.~\cite{micropages} propose a mechanism that identifies
frequently referenced sub-rows of data, and migrates them to reserved
rows.
By co-locating these frequently accessed sub-rows, this scheme aims to
increase the row buffer hit rate of memory accesses, and improve
performance and energy consumption.
DRAM-aware last-level cache writeback schemes~\cite{vwq-isca10,
 lee2010dram} speculatively issue writeback
requests that are predicted to hit in the row buffer.  RBLA is complementary to
these works, and can be applied together \changesI{with them} because RBLA targets a different
problem.

Row buffer locality is also commonly exploited in memory scheduling algorithms.
The First-Ready First-Come-First-Serve algorithm (FR-FCFS)~\cite{frfcfs, frfcfspatent} 
prioritizes memory requests that hit
in the row buffer, improving the latency, throughput, and energy cost of
serving memory requests. 
Many other memory scheduling algorithms~\cite{stfm, parbs, atlas, tcm,pa-micro08,pam,morse-hpca12,lee2010dram,bliss,mise,usui-squash,jishen-firm,ipek-isca2008,cjlee-micro09,vwq-isca10,mutlu-podc08,ghose2013,xiong-taco16,liu-ipccc16,bliss-tpds,lavanya-asm,memattack,mcp,hyoseung-rtas14,hyoseung-rts16,sms,jeong2012qos,usui-dash,fst,ebrahimi-isca2011}
build upon this ``row-hit first'' principle.

Muralidhara et al.~\cite{mcp} use a thread's row
buffer locality as a metric to decide which channel the thread's pages
should be allocated to in a multi-channel memory system. Their goal
is to reduce memory interference between threads, and as such their
technique is complementary to ours.





\section{Significance}

Our ICCD 2012 paper~\cite{rbla} makes several novel contributions that
we expect will have a long-term impact on the design of memory systems,
and we believe that our work inspires several new research questions.

\subsection{Long-Term Impact}

The memory scaling bottleneck continues to be a significant hurdle to system
performance \changesI{and energy efficiency~\cite{mutlu.imw13,superfri,mutlu2017rowhammer}}.  Emerging applications operate on increasingly-larger
sets of data, and require high-capacity, high-performance main memories,
but the poor scaling of DRAM limits the ability of these applications to fit 
their entire working sets within a DRAM-based main memory.  Because DRAM cannot keep pace with
application needs, we expect that the demand for
alternative memory technologies will continue to grow in the coming years.

Hybrid memory systems can allow systems to harness these alternative memory
technologies without fully sacrificing the benefits of DRAM.  By combining slower
but larger memories (e.g., NVM) with faster but smaller memories (e.g., DRAM), 
a hybrid memory system
has the potential to provide the illusion of a fast and large memory system at
a reasonable cost.
However, as we discuss, this potential can only be realized by carefully 
considering which pieces of data are placed in each of the constituent memories
of a hybrid memory system.
To our knowledge, our ICCD 2012 paper~\cite{rbla} is the first to show that
the organization of the underlying memory technologies, such as the existence of
row buffers, can be used to make more intelligent data placement decisions.

While our ICCD 2012 paper~\cite{rbla} shows the impact of our proposed data
placement policy on a hybrid memory consisting of DRAM and PCM, it can be used
to enable a wide range of hybrid memory systems.  For example, STT-MRAM 
devices can make use of a row buffer~\cite{Architect_STTRAM, meza.iccd12, meza.tr12, andre.jssc05},
and expensive reduced-latency DRAM devices~\cite{micron-rldram3, sato-vlsic1998} also make use of a row buffer.
RBLA can be used to improve the performance of hybrid memories that include any of these memory
technologies, as our general observations on row buffer locality remain the 
same.
We expect that this versatility will increase the potential impact of RBLA,
as no single memory technology has yet to emerge as the dominant replacement
for DRAM.

\subsection{Research Questions}

As we show in our ICCD 2012 paper~\cite{rbla}, the efficient management of
hybrid memory systems requires the identification and consideration of the
key similarities and trade-offs of each memory type.
An open research question inspired by RBLA's use of row buffer locality is
\emph{what other properties of memory systems should hybrid memory management
mechanisms consider?}
For example, one of our recent works~\cite{yang-ubm} incorporates information
on memory-level parallelism (MLP) into data placement decisions in
hybrid memory management.  As that work shows, we can use a combination of
access frequency, row buffer locality, and MLP to predict the overall performance 
impact of migrating a page between each memory type.
As future memory technologies are developed, we expect that other such properties
will be important to consider, in order to maximize the benefits provided by the
hybrid memory system.

Several works propose on-chip DRAM caches~\cite{chop, CAMEO, yu.micro17, yu.arxiv17}, where a small amount of DRAM
is used as a last-level cache to reduce the number of accesses to a larger off-chip
DRAM.  This is akin to the design of a hybrid memory system, but there are 
different trade-offs in each design.  For example, while the row buffer hit latency is
typically similar across memory technologies in hybrid memories,
both a row buffer hit and a row buffer miss take longer when accessing the
off-chip DRAM as opposed to accessing the on-chip DRAM cache.
This inspires us to ask \emph{how can principles of hybrid memory systems be
applied to DRAM cache management, and vice versa?}
Extending upon this, \emph{can we design general mechanisms that can be
applied to both hybrid memory systems and DRAM cache management?}
As one example, our recent work~\cite{yang-ubm} on predicting the utility of
data placement decisions is highly parameterized, and these parameters
can easily be tuned to represent the trade-offs in both hybrid memory systems
and in systems with a DRAM cache.
We believe and hope that future works should strive to develop other such general
mechanisms.

\section{Conclusion} 

Our ICCD 2012 paper~\cite{rbla} observes that row buffer access latency (and energy) in
DRAM and PCM are similar, while PCM array access latency (and energy)
is much higher than DRAM array access latency (and energy). Therefore,
in a hybrid memory system where DRAM is used as a cache to PCM, it
makes sense to place in DRAM data that would cause frequent row buffer
misses as such data, if placed in PCM, would incur the high PCM array
access latency. We develop a caching policy that achieves this effect
by keeping track of rows that have high row buffer miss counts (i.e.,
low row buffer locality, but high reuse) and places only such rows in
DRAM. Our final policy dynamically determines the threshold used to
decide whether a row has low locality based on cost-benefit
analysis. Evaluations show that the proposed row buffer locality aware
caching policy provides better performance, fairness, and
energy-efficiency compared to caching policies that only consider
access frequency or recency. Our mechanisms are applicable to 
and can improve the performance of other
hybrid memory systems consisting of different technologies.
We hope that our findings can help ease the adoption of
emerging memory technologies in future systems, and
inspire further research in data management policies.

\section*{Acknowledgments}

We thank Saugata Ghose for his dedicated effort in the preparation
of this article.
We acknowledge the support of AMD, HP, Intel, Oracle,
and Samsung. This research was partially supported by 
the NSF (grants 0953246 and 1212962), GSRC, Intel URO,
and ISTC on Cloud Computing. HanBin Yoon was partially supported
by the Samsung Scholarship.


{
\bibliographystyle{IEEEtranS}
\bibliography{paper}
}

\end{document}